\documentclass[12pt]{article}
\newcommand{\beq}{\begin{equation}}
\newcommand{\eeq}{\end{equation}}
\newcommand{\bra}{\begin{array}}
\newcommand{\era}{\end{array}}

\newcommand{\al}{\alpha}

\newcommand{\de}{\delta}

\usepackage{latexsym}
\author{Jamila Douari\footnote{jdouari@gmail.com}\\ \\
\small\it Center for Advanced Mathematical Sciences\\\small\it
American University of Beirut\\ \small\it P.O.Box 11-0236, College
Hall \\\small\it Beirut, Lebanon\rm }
\title{Neumann Boundary Conditions from D1-Brane Description at the Presence of Electric Field}
\begin{document}
\maketitle \vspace*{0.5cm} PACS: 11.25.-w, 11.25.Uv, 11.15.Kc \vskip1cm
Keywords: Branes, Dyons, Fluctuations, Boundary Conditions.
\vspace*{1.5cm}
\section*{Abstract}
\hspace{.3in}We study the fluctuations of D1$\bot$D3 branes from D1-Brane description in the presence of world volume electric field. The fluctuations are found to obey Neumann boundary conditions separating the system into two regions depending on electric field $E$.
\section{Introduction}
\hspace{.3in} The D-brane's world volume is governed by the Born-Infeld (BI) action which is the dimensional
reduction of 10-dimensional supersymmetric Born-Infeld electrodynamics \cite{BI}. Among the many fascinating features of D-branes there is the possibility for D-branes to morph into other D-branes of different dimensions by
exciting some of the scalar fields \cite{InterBran1,InterBran2}.
It's known in the literature that there are many different but
physically equivalent descriptions of how a D1-brane may end on a
D3-brane. From the point of view of the D3 brane the configuration
is described by a monopole on its world volume. From the point of
view of the D1-brane the configuration is described by D1 opening up
into a D3-brane where the extra three dimensions form a fuzzy two
sphere whose radius diverges at the origin of the D3-brane. These
different view points are the stringy realization of the Nahm
transformation \cite{fun}. The dynamics of both bion spike
\cite{InterBran1,fluct} and fuzzy funnel \cite{dual,fuzfun} were
studied by considering linearized
fluctuations around the static solutions.\\

In this context, the main sections of the present paper will be
devoted to study the fluctuations of funnel solutions using D1-brane theory by considering
dyonic strings. We start by a brief review on D1$\bot$D3 branes in
dyonic case described by the abelian and the non-abelian BI action in section 2. It's known that the description of intersecting fundamental string/D3-branes was presented by Callan and Maldacena \cite{InterBran1} by showing that BI action can be used to bluilt a configuration of a semi-infinite fundamental string ending on a 3-brane. In \cite{NBCBI} it wa showed that by exciting the static solutions of F-string/D3-brane system plus the excitation of electromagnetic field the system can obey Neumann boundary conditions such that the BI dynamics of the F-string was considered. This is looking as a dual description of the system we treat in our present work in which the wave excitation is coming up the dyonic string to D3-brane by using non-abelian BI action, only that we don't assume the fluctuation at the level of electromagnetic field. Thus, in section 3, we discuss the fluctuations of the fuzzy funnel from D1-brane point of view. We give the solutions of the linearized equations of motion of the
fluctuations for overall transverse case. We also discuss the
solutions and the potential depending on the presence of electric
field which is leading to divide the system into two regions meaning that the end of the string moves freely on the D3-brane realizing Polchinski's open string Neumann boundary condition. In section 4, we conclude.

\section{D1$\bot$D3 Branes in Dyonic Case}
\hspace{.3in}In this section, we review in brief the funnel
solutions of D1$\bot$D3 branes from D3 and D1 branes points of
view using abelian and non-abelian BI actions respectively for the world volume gauge
field in dyonic case. The dyonic system is given by considering the D-strings
and the fundamental strings which are introduced by
adding a $U(1)$ electric field. It was showed in \cite{InterBran1,cm} that the BI action, when
taken as the fundamental action, can be used to build a
configuration with a semi-infinite fundamental string ending on a
D3-brane \cite{Gib}. Thus the system is described by the
following action \beq\bra{lll} S=\int dt L &=-T_3 \int
d^4\sigma\sqrt{-det(\eta_{ab}+\lambda^2
\partial_a \phi^i \partial_b \phi^i +\lambda F_{ab})}\\\\
&= -T_3 \int d^4\sigma\Big[ 1 +\lambda^2 \Big( \mid \nabla\phi \mid^2 +B^2 +E^2 \Big)\\\\
&+\lambda^4 \Big( (B.\nabla\phi )^2 +(E.B)^2 +\mid E
\wedge\nabla\phi\mid^2 \Big) \Big]^{\frac{1}{2}} \era\eeq in which
$F_{ab}$ ($a,b=0,...,3$) is the field strength and the electric field is denoted as
$F_{0d}=E$ ($d=1,2,3$). $\sigma^a$ denote the world volume coordinates while $\phi^i$
($i=4,...,9$) are the scalars describing transverse fluctuations of
the brane and $\lambda=2\pi \ell_s^2$ with $\ell_s$ is the string
length. In our case we excite just one scalar so $\phi^i=\phi^9
\equiv\phi$. Considering static gauge, the system gets the lowest energy
at specific condition as we see below. Accordingly to (1) the energy of dyonic system is
\beq\bra{ll} \Xi&= T_3 \int d^3\sigma\Big[ \lambda^2 \mid
\nabla\phi +\stackrel{\rightarrow}{B}
+\stackrel{\rightarrow}{E}\mid^2 +(1-\lambda^2
\nabla\phi.\stackrel{\rightarrow}{B})^2-2\lambda^2
\stackrel{\rightarrow}{E}.(\stackrel{\rightarrow}{B}
+\nabla\phi)\\\\
&+ \lambda^4 \Big(
(\stackrel{\rightarrow}{E}.\stackrel{\rightarrow}{B})^2 +\mid
\stackrel{\rightarrow}{E} \Lambda\nabla\phi\mid^2\Big)
\Big]^{1/2}.\era\eeq Then if we require $\nabla\phi
+\stackrel{\rightarrow}{B} +\stackrel{\rightarrow}{E}=0$, $\Xi$
reduces to the lower bound $\Xi_0\geq0$ \beq\bra{ll} \Xi_0 &=T_3 \int
d^3\sigma\Big[(1-\lambda^2
(\nabla\phi).\stackrel{\rightarrow}{B})^2+2\lambda^2
\stackrel{\rightarrow}{E}.\stackrel{\rightarrow}{E} )\\\\&+
\lambda^4 ( (\stackrel{\rightarrow}{E}.\stackrel{\rightarrow}{B})^2
+{\mid\stackrel{\rightarrow}{E}}\Lambda \nabla\phi\mid^2)
\Big]^{1/2}.\era\eeq By using the Bianchi identity
$ \nabla.B= 0$ and the fact that the gauge field is static, the
funnel solution is then \beq\phi =\frac{N_m +N_e}{2r},\eeq
with $N_m$ is magnetic charge, $N_e$ electric charge and $r=\sqrt{\sigma_1^2 +\sigma_2^2 +\sigma_3^2 }$.\\

In the dual description of the $D1\bot D3$ system we consider ($N, N_f$)-strings; we have
$N$ D-strings and $N_f$ fundamental strings \cite{9911136}. The
theory is described by the action \beq S=-T_1\int d^2\sigma STr
\Big[ -det(\eta_{ab}+\lambda^2 \partial_a \phi^i
Q_{ij}^{-1}\partial_b \phi^j +\lambda F_{ab})det
Q^{ij}\Big]^{1\over2}\eeq in which we replaced the strength field
$F_{\tau\sigma}$ by $EI_{N}$ ($I_{N}$ is $N\times N$-matrix), $
Q_{ij}=\de_{ij} +i\lambda \lbrack \phi_i , \phi_j \rbrack$ and $\phi^i$, $i=1,2,3$, are the scalar fields.\\

The action can be rewritten as \beq S=-T_1\int d^2\sigma STr \Big[
-det\pmatrix{\eta_{ab}+\lambda F_{ab}& \lambda \partial_a \phi^j
\cr -\lambda \partial_b \phi^i & Q^{ij}\cr}\Big]^{1\over2}.\eeq Then
the bound states of D-strings and fundamental strings are made
simply by introducing a background $U(1)$ electric field on
D-strings. By computing the determinant, the action becomes \beq
S=-T_1\int d^2\sigma STr \Big[ (1-\lambda^2 E^2 + \al_i \al_i
\hat{R}'^2)(1+4\lambda^2 \al_j \al_j \hat{R}^4 )\Big]^{1\over2},\eeq
where the following ansatz was inserted \beq\phi_i =\hat{R}\al_i
.\eeq Hence, we get the funnel solution for dyonic string by solving
the equation of variation of $\hat{R}$, as follows \beq \phi_i
=\frac{\al_i}{2\sigma\sqrt{1-\lambda^2 E^2}} .\eeq

\section{Fluctuations of Dyonic Funnel Solutions}
\hspace{.3in}To treat the dynamics of the dyonic funnel solutions in D1-branes description we solve the linearized equations of motion for small and
time-dependent fluctuations of the transverse scalar around the exact background. We deal with the fluctuations of the funnels (9) discussed in section 2. By plugging into the full ($N-N_f$) string action
(6,7) the "overall transverse" $\delta \phi^m (\sigma,t)=f^m
(\sigma,t)I_N$, $m=4,...,8$ which is the simplest type of
fluctuation with $I_N$ the identity matrix; those don't excite
internal modes on $S^2$, together with the funnel solution, we get
\beq\bra{llll} S&=-T_1\int d^2\sigma STr \Big[ (1+\lambda
E)(1+\frac{\lambda^2 \al^i \al^i}{4\sigma^4 }) \Big(
(1+\frac{\lambda^2 \al^i \al^i}{4\sigma^4})(1+(\lambda E
-1)\lambda^2 (\partial_t \delta\phi^m )^2) +\lambda^2
(\partial_\sigma \delta\phi^m )^2\Big)
\Big]^{1\over 2}\\\\
&\approx-NT_1\int d^2\sigma H \Big[ (1+\lambda E)-(1-\lambda^2
E^2)\frac{\lambda^2}{2} (\dot{f}^m)^2 +\frac{(1+\lambda
E)\lambda^2}{2H} (\partial_\sigma f^m)^2 +...\Big] \era \eeq where
$$H=1+\frac{\lambda^2 C}{4\sigma^4}$$ and $C=Tr \al^i \al^i$. For
the irreducible $N\times N$ representation $C=N^2 -1$. In the last
line we have only kept the terms quadratic in the fluctuations as
this is sufficient to determine the linearized equations of motion
\beq\Big( (1-\lambda E)(1+\lambda^2\frac{N^2
-1}{4\sigma^4})\partial^{2}_{t}-\partial^{2}_{\sigma}\Big) f^m
=0.\eeq

In the overall case, all the points of the fuzzy funnel move or
fluctuate in the same direction of the dyonic string by an equal
distance $\delta x^m$. First, the funnel solution is $\phi^i
=\frac{1}{2\sqrt{1-\lambda^2 E^2}}\frac{\al^i}{\sigma}$ and we suggegest that for a definite energy (frequency $\omega$) the
fluctuation $f^m$ waves in the direction of $x^m$ are \beq f^m
(\sigma,t)=\Phi(\sigma)e^{-iwt}\delta x^m.\eeq With this ansatz the
equation of motion (11) becomes \beq\Big( -\partial^{2}_{\sigma}-\frac{\lambda^2 C}{4\sigma^4}(1-\lambda E)w^2
\Big) \Phi(\sigma)=(1-\lambda E)w^2\Phi(\sigma).\eeq 
We remark here that the neergy associated to the present fluctuation $\xi=(1-\lambda E)w^2$ will be negative if $E\gg \frac{1}{\lambda}$ corresponding to unphysical phenomena. To avoid the critical case acquired by large electric field we consider in the following $0\leq E \leq \frac{1}{\lambda}.$ Thus, we remark that the equation (13) is an analog one-dimensional
Schr\"odinger equation with
\beq V(\sigma)=w^2(\lambda E-1)\lambda^2\frac{N^2 -1}{4\sigma^4}.\eeq
We notice that, in the unphysical case in which the electric field dominates, the
potential goes to zero at large $\sigma$ and to $\infty$ for small
$\sigma$. Now, in the physical case, if $E\ll 1/\lambda$ we find $V\rightarrow 0$ for large $\sigma$ and $V\rightarrow -\infty$ if
$\sigma\rightarrow 0$ but if $E\approx 1/\lambda$ the potential goes to zero $\forall \sigma$. This can be seen as two separated systems
depending on electric field so we have Neumann boundary condition
separating the system into two regions $E\approx 0$ and $E\approx 1/\lambda$.\\

Then, the problem is reduced to finding the solution of a single scalar equation. Thus
the first solution we could find is \beq\Phi(\sigma)=e^{\pm (
(1-\lambda E)\frac{w^2}{2}(\sigma^2 +\frac{\lambda^2 (N^2
-1)}{12\sigma^2}) +cst)}.\eeq and the fluctuation then is \beq
f^m(\sigma)=e^{\pm ( (1-\lambda E)\frac{w^2}{2}(\sigma^2
+\frac{\lambda^2 (N^2 -1)}{12\sigma^2}) +cst)}e^{-iwt}\delta
x^m.\eeq These solutions (14) and (15) are taken by considering the
following condition; $$\Big( \partial_\sigma (\sigma^2
+\frac{\lambda^2 (N^2 -1)}{12\sigma^2})\Big)^2 =0,$$ which means $\sigma$ is fixed to be
\beq \sigma=\pm \sqrt{\lambda}(\frac{N^2 -1}{12})^{\frac{1}{4}}.
\eeq The fluctuation (15) is then found on fixed point (16)
depending only on electric field with $w$ ,$\lambda$ and large $N$
fixed
$$f^m(\sigma)=e^{\pm ( (1-\lambda E)\frac{w^2 \lambda N}{2\sqrt{3}}
+cst)}e^{-iwt}\delta x^m$$
and for other points on the string the solution will be different as we will see below.\\

The equation (13) can be also rewritten as follows \beq\Big(
\frac{1}{w^2(1-\lambda E)}\partial^{2}_{\sigma}+1+\frac{\lambda^2
N^2 }{4\sigma^4}\Big) \Phi(\sigma)=0,\eeq for large $N$. Now by
suggesting $\tilde{\sigma}=w\sqrt{1-\lambda E}\sigma$ the latter
equation becomes \beq\Big(
\partial^{2}_{\tilde{\sigma}}+1+\frac{\kappa^2}{\tilde{\sigma}^4}\Big)
\Phi(\tilde{\sigma})=0,\eeq with the potential is
\beq V(\tilde{\sigma})=\frac{\kappa^2}{\tilde{\sigma}^4}, \eeq and
$\kappa=\frac{\lambda N w^2}{2}(1-\lambda E)$, with $E\ne
\frac{1}{\lambda}$. This equation is a Schr\"odinger equation for an
attractive singular potential $\propto\tilde{\sigma}^{-4}$ and
depends on the single coupling parameter $\kappa$ with constant
positive Schr\"odinger energy. The solution is then known by making
the following coordinate change \beq
\chi(\tilde{\sigma})=\int\limits^{\tilde{\sigma}}_{\sqrt{\kappa}}
dy\sqrt{1+\frac{\kappa^2}{y^4}}, \eeq and \beq
\Phi=(1+\frac{\kappa^2}{\tilde{\sigma}^4})^{-\frac{1}{4}}\tilde{\Phi}.
\eeq Thus, the equation (19) becomes \beq\Big(
-\partial^{2}_{\chi}+V(\chi)\Big) \tilde{\Phi}=\tilde{\Phi},\eeq
with \beq V(\chi)=\frac{5\kappa^2}{(\tilde{\sigma}^2
+\frac{\kappa^2}{\tilde{\sigma}^2})^3}. \eeq Then \beq
f^m=(1+\frac{\kappa^2}{\tilde{\sigma}^4})^{-\frac{1}{4}}e^{\pm
i\chi(\tilde{\sigma})}e^{-iwt}\delta x^m . \eeq This fluctuation has
the following limits; at large $\sigma$, $f^m\sim e^{\pm
i\chi(\tilde{\sigma})}e^{-iwt}\delta x^m $ and if $\sigma$ is small
$f^m=\frac{\sqrt{\kappa}}{\tilde{\sigma}}e^{\pm
i\chi(\tilde{\sigma})}e^{-iwt}\delta x^m$. These are the asymptotic
wave function in the regions $\chi\rightarrow \pm\infty$, while
around $\chi\ll 1/\lambda$; i.e. $\tilde{\sigma}\sim\sqrt{\kappa}$,
$f^m\sim 2^{-\frac{1}{4}}e^{-iwt}\delta x^m$.

A closer look at the potential in various limits of electric field we find;\\
\begin{itemize}
\item
$E\ll 1/\lambda$, $V(\chi)\sim \frac{5\lambda^2 N^2 w^2}{4(w^2 \sigma^2
+\frac{\lambda^2 N^2 w^2}{4\sigma^2})}; \phantom{~~}\sigma\sim
\infty\Longrightarrow V(\chi)\sim \infty, \phantom{~~}\sigma\sim
0\Longrightarrow V(\chi)\ll 1/\lambda$ \item $E\sim \frac{1}{\lambda}$,
$V(\chi)\ll 1/\lambda$
\end{itemize}
As discussed above, again we get the separated systems in different
regions depending on the values of electric field. Also if we have a
look at the fluctuation (25) as well as (15) we find that $f^m$ in
the case of $E\approx 1/\lambda$ is different from the one in $E\ll 1/\lambda$ case; as a result the function $f^m$ will get a discontinuity.
This is seen as Neumann boundary condition from non-Born-Infeld dynamics separating the system into two regions depending on the electric field.\\

The fluctuations discussed above could be called the zero mode
$\ell=0$ and for high modes $\ell\geq0$, the fluctuations are
$\delta \phi^m (\sigma,t)=\sum\limits^{N-1}_{\ell=0}\psi^{m}_{i_1
... i_\ell}\al^{i_1} ... \al^{i_\ell} $ with $\psi^{m}_{i_1 ...
i_\ell}$ are completely symmetric and traceless in the lower
indices.\\

The action describing this system is \beq\bra{lll}
S&\approx-NT_1\int d^2\sigma  \Big[ (1+\lambda E)H-(1-\lambda^2
E^2)H\frac{\lambda^2}{2} (\partial_{t}\delta\phi^m)^2) \\\\
&+\frac{(1+\lambda E)\lambda^2}{2} (\partial_\sigma \delta\phi^m)^2
-(1-\lambda^2 E^2)\frac{\lambda^2}{2}\lbrack \phi^i ,\delta\phi^m
\rbrack^2 \\\\
&-\frac{\lambda^4}{12}\lbrack \partial_{\sigma}\phi^i
,\partial_{t}\delta\phi^m \rbrack^2+...\Big] \era.\eeq The
linearized equations of motion are \beq \Big[(1-\lambda
E)H\partial_{t}^2 -\partial_{\sigma}^2\Big]\delta\phi^m
+(1-\lambda E)\lbrack \phi^i ,\lbrack \phi^i ,\delta\phi^m
\rbrack\rbrack -\frac{\lambda^2}{6(1+\lambda E)}\lbrack \partial_{\sigma}\phi^i
,\lbrack \partial_{\sigma}\phi^i ,\partial^2 _{t}\delta\phi^m
\rbrack\rbrack=0.\eeq The backgound solution was $\phi^i \propto
\al^i$ and we have $\lbrack \al^i , \al^j \rbrack
=2i\epsilon_{ijk}\al^k $. Thus \beq\bra{ll} \lbrack \al^i ,
\lbrack\al^i, \delta\phi^m \rbrack
&=\sum\limits_{\ell<N}\psi^{m}_{i_1 ... i_\ell}\lbrack \al^i ,
\lbrack\al^i ,\al^{i_1} ... \al^{i_\ell} \rbrack\\\\
&=\sum\limits_{\ell<N}4\ell(\ell+1)\psi^{m}_{i_1 ...
i_\ell}\al^{i_1} ... \al^{i_\ell} \era\eeq To obtain a specific
spherical harmonic on 2-sphere, we have \beq\lbrack \phi^i ,\lbrack
\phi^i ,\delta\phi_{\ell}^m
\rbrack\rbrack=\frac{\ell(\ell+1)}{\sigma^2}\delta\phi_{\ell}^m
,\phantom{~~~~~~}\lbrack \partial_{\sigma}\phi^i ,\lbrack
\partial_{\sigma}\phi^i ,\partial_{t}^2 \delta\phi^m
\rbrack\rbrack=\frac{\ell(\ell+1)}{\sigma^4}\partial_{t}^2\delta\phi
_{\ell}^m .\eeq Then for each mode the equations of motion are \beq
\Big[ -\partial_{\sigma}^2 -w^2 \Big((1-\lambda E)(1+\lambda^2\frac{N^2 -1}{4\sigma^4})
-\frac{\lambda^2\ell(\ell+1)}{6(1+\lambda E)\sigma^4}\Big)+(1-\lambda E)\frac{\ell(\ell+1)}{\sigma^2}
\Big]\delta\phi_{\ell}^m =0.\eeq The solution of the equation of
motion can be found by taking the following proposal. Let's consider
$\delta\phi_{\ell}^m =f^m_\ell (\sigma)e^{-iwt}$ in direction $m$ with
$f^m_\ell (\sigma)$ is some function of $\sigma$ for each mode
$\ell$.

The last equation can be rewritten as \beq \Big[-\partial_{\sigma}^2
+V(\sigma) \Big] f_{\ell}^m (\sigma)= w^2(1-\lambda E)f_{\ell}^m (\sigma),\eeq with the potential $$V(\sigma)=\frac{w^2\lambda^2}{\sigma^4}\Big(\frac{\ell(\ell+1)}{6(1+\lambda E)}-\frac{(1-\lambda E)C}{4}\Big)+(1-\lambda E)\frac{\ell(\ell+1)}{\sigma^2}.$$ For high mode we remark the following; at large $N$
we get
\begin{itemize} \item if $E\sim 1/\lambda$, $V(\sigma)\sim
\frac{w^2\lambda^2\ell(\ell+1)}{12\sigma^4}$;\\\\ $\phantom{~~}\sigma\sim
\infty\Longrightarrow V(\sigma)\ll 1/\lambda, \phantom{~~}\sigma\sim
0\Longrightarrow V(\sigma)\sim \infty$ \item $E\ll 1/\lambda$,
$V(\sigma)\sim \frac{w^2\lambda^2}{\sigma^4}\Big(\frac{\ell(\ell+1)}{6}-\frac{C}{4}\Big)+\frac{\ell(\ell+1)}{\sigma^2};\\\\ \phantom{~~}\sigma\sim
\infty\Longrightarrow V(\sigma)\ll 1/\lambda, \phantom{~~}\sigma\ll 1/\lambda\Longrightarrow V(\sigma)\sim -\infty$.
\end{itemize}

The equation (30) can be rewritten as 
\beq
\Big[\partial_{\tilde{\sigma}}^2 + 1+
\frac{\tilde{\kappa}^2}{\tilde{\sigma}^4}+\frac{\eta}{\tilde{\sigma}^2}
\Big] f_{\ell}^m (\sigma)= 0,\eeq
in which we defined a new coordinate $\tilde{\sigma}=w\sqrt{1-\lambda E}\sigma$, ($0\leq E< 1/\lambda$) and 
$$
\tilde{\kappa}^2=w^4\lambda^2(1-\lambda E)^2\Big(\frac{N^2}{4}-\frac{\ell(\ell+1)}{6(1-\lambda^2 E^2)}\Big),\phantom{~~~~} \eta=-(1-\lambda E)\frac{\ell(\ell+1)}{w^2}
$$
such that $$N>\sqrt{\frac{2\ell(\ell+1)}{3(1-\lambda^2 E^2)}}.$$ 
To solve the above equation we suggest the following steps. First, we choose small $\sigma$, the equation (32) is
reduced to \beq \Big[\partial_{\tilde{\sigma}}^2 + 1+\frac{\tilde{\kappa}^2}{\tilde{\sigma}^4} \Big] f_{\ell}^m (\sigma)= 0.\eeq
As we saw in mode zero, this kid of equation can be solved by using WKB approach. Thus, to get the solution we use the steps
(21-25). Since we considered small $\sigma$ the potential as found in (25) is evaluated to be
$$
V(\chi)=\frac{5\tilde{\sigma}^6}{\tilde{\kappa}^4}.
$$
for $\chi$ defined by (21) for $\tilde{\kappa}$ instead of $\kappa$.
Then, the solution of (33) takes the form \beq f^m_{\ell}=\frac{\tilde{\sigma}}{\sqrt{\tilde{\kappa}}}e^{\pm
i\chi(\tilde{\sigma})}e^{-iwt}\delta x^m . \eeq 
This fluctuation will get two different values at different limits of $E$ leading to its discontinuity on the string. A closer look at
the potential at large and fixed $N$ in the limits of electric field leads to
\begin{itemize}
\item $E\ll 1/\lambda$, $V(\chi)\sim\frac{5\tilde{\sigma}^6}{w^2 \lambda^4 \Big(\frac{N^2
}{4}-\frac{\ell(\ell+1)}{6}\Big)^2}$,
\item $E\sim 1/\lambda$, $V(\chi)\sim \frac{-60\tilde{\sigma}^6}{w^2 \lambda^4 \ell(\ell+1)}$.
\end{itemize}
Thus, by considering just a small $\sigma$ we see that the system is separated to two regions. Now, let's check the case of large $\sigma$. The
equation of motion (32) of the fluctuation takes the following form by going back to the original variable $\sigma$ \beq
\Big[-\partial_{\sigma}^2 +\tilde{V}(\sigma) \Big] f_{\ell}^m
(\sigma)= w^2(1-\lambda E)f_{\ell}^m(\sigma),\eeq where $$\tilde{V}(\sigma)= \frac{(1-\lambda E)\ell(\ell+1)}{\sigma^2}$$ and $f_{\ell}^m$
is now a Sturm-Liouville eigenvalue problem. The solution then is as follows
\beq
f_{\ell}^m(\sigma)=\sqrt{\sigma}\Big( \alpha J(a,b)+\beta Y(a,b) \Big)
\eeq
$$a=\frac{\sqrt{1+4(1-\lambda E)\ell(\ell+1)}}{2}, \phantom{~~~~}b= w\sqrt{1-\lambda E}$$
with $J$ and $Y$ are Bessel functions of first and second kind respectively and $\alpha$, $\beta$ are some constants.\\

Concerning the potential we remak the following in the different regions of electric field; if $E\sim 1/\lambda$ the potential is zero for all large $\sigma$ and if $E\ll 1/\lambda$ the potential is going to zero. In this last case we don't see big difference by taking the limits for $E$ but by connecting the results discussed for small $\sigma$ to the large one we get clearly different values for the potental in the two cases $E\sim 1/\lambda$ and $E\ll 1/\lambda$. Consequently, the system is divided to two regions meaning that having Neumann boundary conditions and the end of the string can move freely on the brane.

\section{Discussion and Conclusion}
\hspace{.3in}The description of intersecting fundamental string/D3-brane was presented by Callan and Maldacena \cite{InterBran1} by showing that BI action can be used to build a configuration of a semi-infinite fundametal string ending on a 3-brane. They also showed that the fluctuations which are normal to both the string and the 3-brane behave as if they had Dirichlet boundary conditions at the point of attachement. But as we see in the reference \cite{NBCBI} in which the fluctuations were at the level of static solution and electromagnetic field, the excitations are coming down the string with a polarization along a direction parallel to the brane are almost completely reflected but the end of the string moves freely on the 3-brane realizing Neumann boundary condition dynamically. In our work, we consider the dual description by exciting only the BPS solution of the system constituting of dyonic strings (instead of just fundamental or magnetic strings) and D3-brane. We found that the fluctuations behave like having Neumann boundary condition at the presence of electric field $E$ in the system such that $0\leq E\leq 1/\lambda$.\\

In this work we reviewed in brief the funnel solutions
for D1$\bot$D3 branes from D3 and D1 branes points of view and we
studied the dynamics of the funnel solutions by plugging into the
full ($N -N_{f}$ ) string action the "overall transverse"
fluctuations. We discussed the fluctuations and the potentials in
the zero and high modes of the overall transverse. In both cases, we
found that D1$\bot$D3 system has Neumann boundary conditions caused
by the presence of electric field. When the electric field is going
up and down the potential of the system and the fluctuations are
changing. Consequently, the end point of the dyonic string moves on
the brane.

It would be interesting to see whether the properties found in this
paper would also be shown in the relative transverse case and for
high dimensional branes. Also, we could deal with supergravity
background and see if we will get the same boundary conditions by
treating the dyonic fluctuations.

\section{Acknowledgements}
\hspace{.3in}This work was supported by a grant from the Arab Fund
for Economic and Social Development. The author would like to thank
Robert de Mello Koch for very pleasant and helpful discussion and
Stellenbosch Institute for Advanced Study where this work was
started.


\begin{thebibliography}{99}
\bibitem{BI} J. Polchinski, Tasi Lectures on D-branes,
hep-th/9611050; R. Leigh, Mod. Phys. Lett .A4 (1989) 2767.

\bibitem{InterBran1} C.G. Callan and J.M. Maldacena, Nucl. Phys.
B513 (1998) 198, hep-th/9708147.

\bibitem{InterBran2} G.W. Gibbons, Nucl. Phys. B514
(1998) 603, hep-th/9709027; P.S. Howe, N.D. Lambert and P.C. West,
Nucl. Phys. B515(1998) 203, hep-th/9709014; T. Banks, W. Fischler,
S. H. Shenker and L. Susskind, Phys. Rev. D55 (1997) 5112,
hepth/9610043; D. Kabat and W. Taylor, Adv. Theor. Math. Phys. 2
(1998) 181, hep-th/9711078; S. Rey, hep-th/9711081; R.C. Myers, JHEP
9912 (1999) 022, hep-th/9910053.

\bibitem{fun} D. Brecher, Phys. Lett. B 442 (1998)
117, hep-th/9804180; P. Cook, R. de Mello Koch and J. Murugan, Phys.
Rev. D 68 (2003) 126007, hep-th/0306250; N. R. Constable, R. C.
Myers and O. Tafjord, Phys. Rev. D 61 (2000) 106009, hep-th/9911136.
J. K. Barrett and P. Bowcock, hep-th/0402163.

\bibitem{fluct} S.-J. Rey and J.-T. Yee, Nucl. Phys. B52 (1998) 229, hep-th/9711202; S. Lee, A. Peet and L.
Thorlacius, Nucl. Phys. B514 (1998) 161, hep-th/9710097; D. Kastor
and J. Traschen, Phys. Rev. D61 (2000) 024034, hep-th/9906237; S.-J.
Rey and J.-T. Yee, Eur. Phys. J. C22 (2001) 379, hep-th/9803001

\bibitem{dual}N. R. Constable, R. C. Myers and O. Tafjord, Phys. Rev. D61, 106009 (2000),
hepth/9911136.

\bibitem{fuzfun} N. R. Constable, R.C. Myers and O. Tafjord, Phys.
Rev. D61, 106009 (2000), hepth/9911136; R. Bhattacharyya and R. de
Mello Koch, hep-th/0508131; C. Papageorgakis, S. Ramgoolam, N.
Toumbas, JHEP 0601 (2006) 030, hep-th/0510144.


\bibitem{NBCBI}K. G. Savvidy and G. K. Savvidy, Nucl. Phys. B561 (1999) 117,
hep-th/9902023.

\bibitem{cm} N. R. Constable, R. C. Myers, O. Tafjord, JHEP 0106 (2001) 023, hep-th/0102080.

\bibitem{Gib} G. W. Gibbons, Nucl.Phys. B514 (1998) 603-639, hep-th/9709027.

\bibitem{9911136} N. R. Constable, R. C. Myers, O. Tafjord,
Phys.Rev. D61 (2000) 106009, hep-th/9911136.


\end{thebibliography}
\end{document}